\documentstyle[twocolumn,aps,prl,epsf,floats]{revtex}
\newcommand{\be}{\begin{equation}}
\newcommand{\ee}{\end{equation}}
\newcommand{\ba}{\begin{eqnarray}}
\newcommand{\ea}{\end{eqnarray}}
\newcommand{\bi}{\begin{itemize}}
\newcommand{\ei}{\end{itemize}}

\newcommand{\hal}{{1\over 2}}
\def\bx{{\bf x}}

\def\lsi{\raise0.3ex
\hbox{$<$\kern-0.75em\raise-1.1ex\hbox{$\sim$}}}
\def\gsi{\raise0.3ex
\hbox{$>$\kern-0.75em\raise-1.1ex\hbox{$\sim$}}}

\begin{document}
\twocolumn[\hsize\textwidth\columnwidth\hsize\csname
@twocolumnfalse\endcsname

\draft
\title{QCD thermodynamics from 3d adjoint Higgs model}
\author{F. Karsch$^{a}$,
M. Oevers$^{b}$,
A. Patk\'os$^{c}$,
P. Petreczky$^{a,c}$ and 
Zs. Sz\'ep$^{c}$
}
\address{$^{a}$Fakult\"at f\"ur Physik, Universit\"at Bielefeld,\\
\qquad\qquad  P.O. Box 100131, D-33501 Bielefeld, Germany} 
\address{$^{b}$Department of Physics and Astronomy, University of
Glasgow,\\
\qquad\qquad Glasgow,  G12 8QQ, U.K. }
\address{$^{c}$Dept. of Atomic Physics, E\"otv\"os University,\\
\qquad H-1088 Puskin 5-7, Budapest Hungary}
\date{\today}
\maketitle

\vspace*{-5.0cm}
\noindent
\hfill \mbox{BI-TP 98/26}
\vspace*{4.8cm}

\begin{abstract}\noindent
Screening masses of hot $SU(N)$ gauge theory, defined as poles of
the corresponding propagators are studied in 3d adjoint
Higgs model, considered as an effective theory of $QCD$,
using coupled gap equations 
and lattice Monte-Carlo simulations (for $N=2$).  In so-called 
$\lambda$ gauges non-perturbative evidence is given for the 
gauge independence of pole masses within this class of gauges.
Application of screening masses in a novel resummation prescription 
of the free energy density is discussed.
\end{abstract}

\vspace*{0.2cm}

\vskip1.5pc]

\section{Intoduction} 
Finite temperature $SU(N)$ theory undergoes a phase transition at some
temperature $T_c$ from the confined to the deconfined phase. Above this 
temperature chromoelectric fields are screened with finite 
inverse screening
length, the inverse of the so-called  Debye mass. Although there is no 
confinement for temperatures above $T_c$, naive perturbation theory is
known to fail because of severe IR divergencies. The by-now well known
resummation techniques, though solve a part of this problem (e.g. a weak
coupling expansion can be obtained up $g^5$ order for the free energy) the
resulting resummed series shows very bad convergence properties
\cite{arnzhai,braat1} \footnote{In \cite{pade} it was shown that
the convergence of the perturbative result for the free energy
density can be improved using Pad`e approximants}. 
For the Debye mass IR problems imply that the naive definition 
$\Pi_L(k_0=0,k \rightarrow 0)$ (where $\Pi_L$
is the longitudinal  self-energy ) is not applicable beyond the leading 
order.  Rebhan has suggested to define the Debye mass as the pole
of the logitudinal part of the propagator \cite{rebhan1}. This definition 
yields gauge invariant results, however, it requires the introduction of 
the so-called magnetic mass, a concept, which was introduced long ago
to cure IR divergencies due to static magnetic fields \cite{linde}.
Analogously to the electric mass the magnetic mass can be defined as
the pole of the transverse gauge boson propagator.
Although the magnetic mass is non-calculable in perturbation theory, 
numerical lattice 
studies of the finite temperature gluon propagator indicate its existence 
\cite{heller1,heller2}.       
Also a self-consistent resummation of perturbative series in 3d  gauge 
theories, considered as an effective theory yields a non-vanishing magnetic 
mass \cite{buchm1,alex,jackiw,eberlein}. 
The 3d effective theory emerges from the
high temperature limit of $SU(N)$ gauge theory as follows: If the temperature
is high enough the asymptotic freedom ensures the separation of
different mass scales $2 \pi T \gg g T \gg g^2 T$ and one can  integrate 
out the heavy modes, with wave numbers $\sim 2 \pi T$ and $\sim g T$.
This yields effective theories describing IR dynamics at scale
$g T$ (adjoint Higgs model) and $g^2 T$ (3d pure gauge theory), respectively
\cite{eff3d}. The usefulness of the effective theory approach for unrevealing
the source of breakdown of perturbation theory and solving the problem of
the perturbative IR catastrophe was illustrated in \cite{braat1}. The effective
theory approach was used to study non-perturbative correction to the
Debye screening mass \cite{arnyaff,kajantie1,kajantie2,laine1}. The mass gap
of pure 3d gauge theory  can be related to the magnetic mass of hot $SU(N)$
gauge theory  through the dimensional reduction.
The presence of the mass gap in 3d gauge theory needs some
comments. In \cite{arnyaff} it has been claimed that the magnetic mass
cannot exist in a confining theory, since it would imply a perimeter
law for large Wilson loops. However, this argument is valid only at tree
level and non-perturbative analysis of the 2+1 dimensional gauge theory 
\cite{karab1,karab2} shows that both magnetic mass and
non-zero string tension are present in the theory at the same time.

Although progress has been made in understanding the high 
temperature dynamics in the case of $QCD$
the usefullness of the effective theory approach is questionable,
because the coupling constant is large for any realistic temperature and 
therefore the separation of different length scales is not obvious.

In this contribution  we will try to clarify whether screening masses
defined as poles of the corresponding propagators
can be determined in the 3d adjoint Higgs model, considered as an effective
theory of hot $SU(N)$ theory (e.g. $QCD$). We are going to review recent papers
of the authors \cite{patkosk,karschk}. The investigation of the 
screening masses defined as poles of the corresponding propagators is of great
interest because
they present essential input into the perturbative calculation of the thermodynamical
quantities \cite{arnzhai,braat1}. 
In \cite{karsch3d} it was shown that the contribution of the
magnetostatic sector to the free energy density can be understood in terms 
of massive quasiparticles with mass $\sim g^2 T$. The outcome of the
resummed perturbative calculation for the free energy may  depend on the 
correct choice of the screening mass \cite{karschsk,landshoff}. 
Therefore we will also discuss
how far the values of the screening masses determined here influence the
result of the perturbative free energy calculation.
We will proceed as follows:
in section 2 coupled gap equations for the screening masses are introduced
and analyzed numerically. In section 3 the lattice formulation of the
adjoint Higgs model is considered,
numerical results for $SU(2)$ Landau gauge propagators are discussed
and compared with recent results from 4d simulation of hot $SU(2)$ theory.
In section 4 propagators
are studied in the so-called $\lambda$-gauges \cite{bernard} and numerical
arguments for gauge independence of the pole mass within this class
of gauges will be presented.
In section 5 application of the pole masses to the resummation of the
free energy is discussed. 

\section{Coupled gap equation}
As was discussed in the introduction a gauge invariant definition 
of screening masses through
the pole of the gluon propagator is possible and they can  be determined
self-consistently through the gap equations 
\cite{rebhan1,buchm1,alex,jackiw,eberlein}. However, 
the determination of the electric and magnetic masses
was attempted independently form each other. In view of the fact that 
$g \sim 1$ it seems natural to determine the screening masses from
a coupled system of gap equations which does not rely on the separation of the
electric and magnetic scales. We will assume, however, the decoupling
of non-static modes and the coupled gap equation will be derived from
the effective theory, the 3d adjoint Higgs model. The lagrangian of
this theory in the Euclidian formulation is the following:
\be
L={1\over 4} F_{ij}^a F_{ij}^a+{1\over 2} {(D_i A_0)}^2+
{1\over 2} m_{D0}^2 (A_0^a A_0^a)+\lambda_A {(A_0^a A_0^a)}^2
\ee
Gauge invariant approaches  for the magnetic mass
generation in three-dimensional pure $SU(N)$ gauge theory  were
suggested by Buchm\"uller and Philipsen (BP) \cite{buchm1} and by
Alexanian and Nair (AN) \cite{alex}.
The approach of AN uses a hard thermal loop inspired effective action
for the resummation in the magnetic sector. The approach of BP makes use of  a
gauged $\sigma$-model, and goes over to the $SU(N)$ gauge theory in 
the limit of
an infinitely heavy scalar field. At present only these two gauge invariant
schemes are known to provide real values for the magnetic mass \cite{jackiw}.
The corresponding expression for the on-shell
self-energy reads
\be
\Pi_T(k=i m_T,m_T)=C m_T,
\ee
where
\be
\label{egy6}
C=\cases{{g^2_3 N\over 8 \pi} [{21\over 4} \ln 3-1],AN ,\cr
{g^2_3 N\over 8 \pi} [{63\over 16} \ln 3-{3\over 4}],BP.}
\ee

Since we are interested in calculating the screening masses in the
three-dimensional $SU(N)$ adjoint Higgs model, $\Pi_T(k,m_T)$ should be
supplemented by the corresponding contribution coming from $A_0$
fields. This contribution is calculated from diagrams:
\vskip0.2truecm
\epsfbox{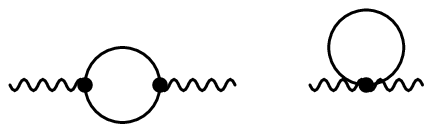}
\noindent
and its analytic expression is
\ba
&&
\delta \Pi_{ij}^{A_0}(k,m_D)=\\
&&
{g^2_3 N\over 4 \pi} \left(-{m_D\over 2}+{k^2+4 m_D^2\over 4 k}
\arctan{k\over 2 m_D}\right) \left(\delta_{ij}-{k_{i} k_{j}\over k^2}\right).
\ea
It is transverse and gauge parameter independent, it also does
not depend on the specific resummation scheme applied to the
magnetostatic sector. It should be also noticed that it starts to contribute to
the gap equation  at ${\cal O}(g^5)$ level in the
weak coupling regime, thus preserving the magnetic mass scale to
be of order $g^2 T$. This is the reason why no "hierarchy" problem arises in
this case in the weak coupling region.
\begin{figure}
\epsfxsize=10cm
\epsfysize=8cm
\epsffile{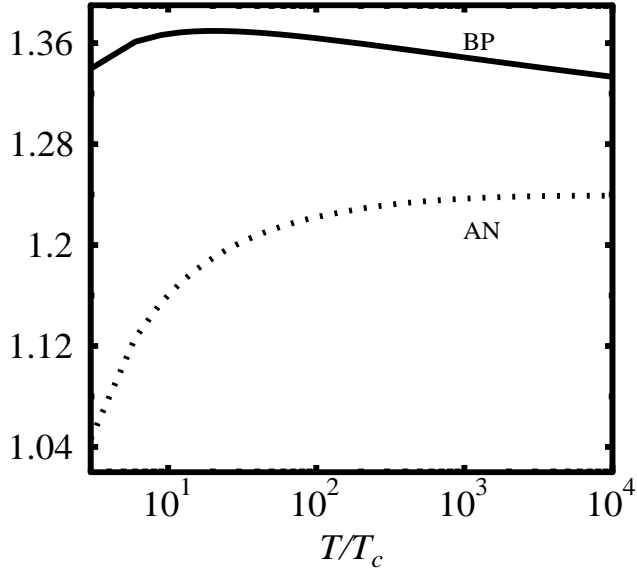}
\caption{The temperature dependence of the scaled Debye mass for
BP resummation scheme (solid) and for the AN resummation scheme
(dashed). The scaling factor is $m_{D0}$.}
\end{figure}

\begin{figure}
\epsfxsize=10cm
\epsfysize=8cm
\centerline{\epsffile{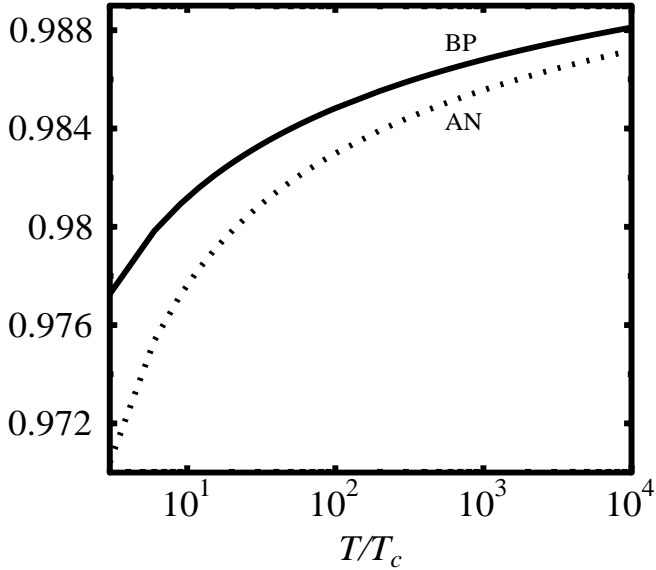}}
\caption{The temperature dependence of the scaled magnetic  mass for
BP resummation scheme (solid) and for the AN resummation scheme
(dashed). The scaling factors are $m_{T}^{BP}$ and $m_{T}^{AN}$,
respectively.}
\end{figure}

At 1-loop order the following diagrams contribute to the $A_0$-self energy
\footnote{There is also a diagram arising from quartic
self coupling of $A_0$, however, since the corresponding coupling is very 
small, we have not taken it into account.}
\vskip0.2truecm
\epsfbox{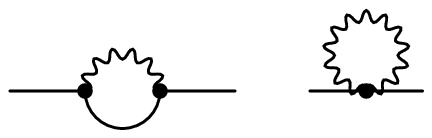}
\noindent

The self energy of $A_0$ depends on the specific resummation scheme.
For BP resummation it reads
\ba
&
\Pi_{00}(k,m_D,m_T)=\nonumber\\
&
{g^2_3 N\over 4 \pi} \biggl[-m_D-m_T+
{2 (m_D^2-k^2-m_T^2/2)\over k} \arctan{k\over
m_T+m_D}-\nonumber\\
&
{(k^2+m_D^2)\over m_T^2} \biggl(-m_T+{(k^2+m_D^2)\over k}
\arctan{k\over m_T+m_D}
\biggr)\biggr].
\ea

In the AN scheme the corresponding expression reads
\ba
&
\Pi_{00}(k,m_D,m_T)=\nonumber\\
&
{g^2_3 N\over 4 \pi} \biggl[-m_D-m_T+
{2 (m_D^2-k^2-m_T^2/2)\over k} \arctan{k\over m_D+m_T}\nonumber\\
&
+(k^2+m_D^2) \biggl({k^2+m_D^2\over m_T^2 k} (\arctan{k\over
m_D+m_T}-\nonumber\\
&
\arctan{k\over m_D+m_T})\biggr) \biggr].
\label{reb}
\ea
Although the value of $\Pi_{00}$ is different for the two resummation scheme
the analytic properties and on-shell values are the same. 
The coupled system of gap equations can be written as
\ba
\nonumber
m_T^2&=&C m_T+\delta \Pi^{A_0}(k=i m_T,m_D), \\
m_D^2&=&m_{D0}^2+\Pi_{00}(k=i m_D,m_D,m_T).
\label{scmasses}
\ea

The temperature dependence of $m_D$ obtained from (\ref{scmasses}) is shown 
in Figure 1 for both schemes, where we have normalized the
Debye mass by the leading order result, $m_{D0}$. The temperature dependence of
the magnetic  mass is shown in Figure 2, where we have normalized $m_T$ by the
value of the magnetic mass obtained for pure three-dimensional $SU(2)$ theory,
in the BP (AN) gauge invariant calculations \cite{buchm1,alex}. As
one can see the influence of $A_0$ 
on the magnetic mass is between 1 and 3\%.
From Figures 1 and 2 it is also seen that the temperature dependence of the
screening masses is very similar to the temperature dependence of the
respective leading order results.

\section{Lattice formulation of 3d adjoint Higgs model and its
physical phase}
The lattice action for the 3d adjoint Higgs model used in the present paper
has the form
\ba
&&
S=\beta \sum_P \hal Tr U_P + \nonumber\\
& &\beta \sum_{\bx,\hat i} \hal Tr A_0(\bx) U_i(\bx) A_0(\bx+\hat i)
U_i^{\dagger}(\bx) + \nonumber\\
&&
\sum_{\bx} \left[-\beta\left(3+\hal h\right) \hal Tr A_0^2(\bx) +
\beta x { \left( \hal Tr
A_0^2(\bx)\right)}^2 \right],\hspace*{-0.4cm}
\label{act}
\ea
where $U_P$ is the plaquette, $U_i$ are the usual link variables and
the adjoint Higgs field is parameterized 
by anti-hermitian matrices $A_0=i \sum_a \sigma^a A_0^a$ ($\sigma^a$ 
are the usual Pauli matricies), as in \cite{kajantie1,kark1}. Furthermore
 $\beta$ is the lattice gauge coupling, $x$ parameterizes the
quartic self coupling of the Higgs field and $h$ denotes the
bare squared mass of the Higgs particle.
This model is known to have two phases the symmetric (confinement)
and the broken (Higgs) phase \cite{nadkarni} separated by the line
of $1^{st}$ order phase transition for small $x$. The strength of the 
transition is decreasing as $x$ increases and turns into a smooth crossover
at $x \sim 0.3$ \cite{kajantie1,hart}. 

The high temperature phase of the
$SU(2)$ gauge theory corresponds to some surface in the parameter
space $(\beta, x, h)$, the surface of 4d physics. This surface may lie in
the symmetric phase or in the broken phase, i.e. the physical phase in
principle might be either the symmetric or the broken phase.
In the previous section it was tacitly assumed that the symmetric phase
is the physical one. This seems to be reasonable because $m_{D0}^2>0$ and at
tree level no symmetry breaking occurs. However, it was shown 
that the 1-loop \cite{polonyi} and the 2-loop \cite{kajantie1} effective
potential of $A_0$ has a non-trivial minimum, i.e. the broken phase
might be the physical. The conclusion that the high-T physics is mapped
onto the broken phase of the 3d model is self-contradictory has been argued
in a recent numerical
lattice investigation, where the parameters appearing in (\ref{act}) were 
determined from a 2-loop
 dimensional reduction \cite{kajantie1}.
The dimensional reduction performed in \cite{kark1} on the other hand
led to the conclusion
that the physical phase is the symmetric one.
The expectation value of the adjoint Higgs field in the broken phase
is of order $\sim {\cal O}({1\over g})$, therefore for $g<<1$
 dimensional reduction is valid only if $g_3 A_0<<T$. What happens 
at intermediate couplings, however, remains an open question.

We will discuss two possibility for fixing the
parameters appearing in (\ref{act}) by
matching non-perturbatively some quantities which are equally 
well calculable both in the full
4d  lattice theory and in the effective 3d lattice theory.
We propose to match  Landau gauge correlators of static link configurations
calculated in the 4d theory to the results of the effective model.
These are quantities calculated in gauges fixed independently.
The assumption is that static 
configurations saturate also the correlators of the full theory.
For such configurations the 4d Landau
gauge is equivalent to its 3d version.

In general this strategy requires the realisation of a matching in a three
dimensional parameter space
($\beta, ~x,~h$), which is clearly a difficult task. We followed
a more modest approach and fixed two of the three parameters,
namely $\beta$
and $x$ at values obtained from the perturbative
dimensional reduction. The values of these parameters
at 2-loop level are \cite{kajantie1}
\vskip0.3truecm
\ba
&&
\beta={4\over g_3^2 a}, \nonumber \\[0.1cm]
&&
g_3^2=g^2\left(\mu\right) T \left[1+{g^2\left(\mu\right)\over 16 \pi^2}
\left(L+{2\over3}\right)\right], \\
&&
x={g^2\left(\mu\right)\over 3 \pi^2}\left[1+
{g^2\left(\mu\right)\over 16 \pi^2} \left(L+4\right)\right],\\
&&
L={44\over 3} \ln{\mu\over 7.0555 T}
\ea
with $a$
and $T$ denoting the lattice spacing and the temperature, respectively.
The coupling constant of the 4d theory $g^2(\mu)$ is defined through
the 2-loop expression
\be
g^{-2}(\mu)={11\over 12 \pi^2} \ln{\mu\over \Lambda_{\overline{MS}}}+
{17\over 44 \pi^2}\ln\left[2\ln{\mu\over
\Lambda_{\overline{MS}}}\right].
\label{4dg}
\ee
For the comparison of the results of the 3d and 4d simulations with
 it is necessary to fix the renormalization and the temperature
scale. We choose the renormalization scale $\mu=2 \pi T$, which ensures
that corrections to the leading order results for the parameters
$g_3^2$ and $x$
of the effective theory are small. Furthermore we use the
relation $T_c=1.06 \Lambda_{\overline{MS}}$ from \cite{heller2}.
Now the temperature scale is fixed
completely and the physical temperature may be varied by varying the
parameter $x$.


Let us discuss the selection of the value of $h$ for fixed $\beta$
(lattice spacing), i.e. the ways how to choose the line of 4d physics.
In our investigations three alternative choices for the line of 4d physics
$h(x)$ have been explored. A comparison with 4d simulations then allows the 
determination of the physical line $h(x)$.
The first alternative for $h(x)$
is the perturbative line of 4d physics, {calculated in \cite{kajantie1}
and} lying in the broken phase,
the other two  are in the symmetric phase.
The three alternatives are illustrated on Figure 3 for $\beta=16$.
The actual procedure we used to  choose the parameter $h$ in the
symmetric phase is the following.
First, we determined the transition line $h_{tr}(x)$. The transition line
as function of $x$ in the infinite volume limit was found in
\cite{kajantie1} in terms of the renormalized mass parameter
$y=m^2/g_3^4$ ($m$ is the continuum renormalized mass). It turns out to be 
independent of $\beta$.
Then using eq. (5.7) from \cite{kajantie1} one can calculate $h_{tr}(x)$.
\begin{figure}
\epsfysize=8cm
\epsfxsize=8cm
\centerline{\epsffile{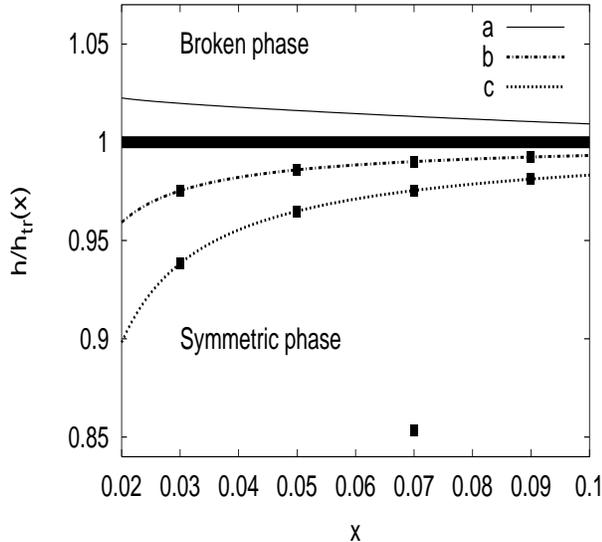}}
\caption{The bare masses normalized by the critical mass $h_{tr}(x)$
used in our matching analysis (squares): the perturbative line (a) and
two alternative sets $I$ (line b) and $II$ (line c). For a discussion of their
choice see text. For $x=0.07$ also a point deeper in the symmetric phase
has been explored.
The thick solid line is the transition line and its
thickness indicates the  uncertainty in its value.}
\end{figure}
The use of the infinite volume result for the transition
line seems to be justified
because most of our simulations were done  on a larger
$32^2 \times 64$ lattice.
The two sets of $h(x)$ values, which appear on Figure 3,
were chosen so that the renormalized mass
parameter $y$ (calculated using eq. (5.7) of \cite{kajantie1} ) always
stays $10 \%$ and $25 \%$ away from the transition line.
These values of $h$ are of course {\em ad hoc} and one should use them only as
trial values.

\begin{figure}[t]
\epsfysize=8cm
\epsfxsize=9cm
\centerline{\epsffile{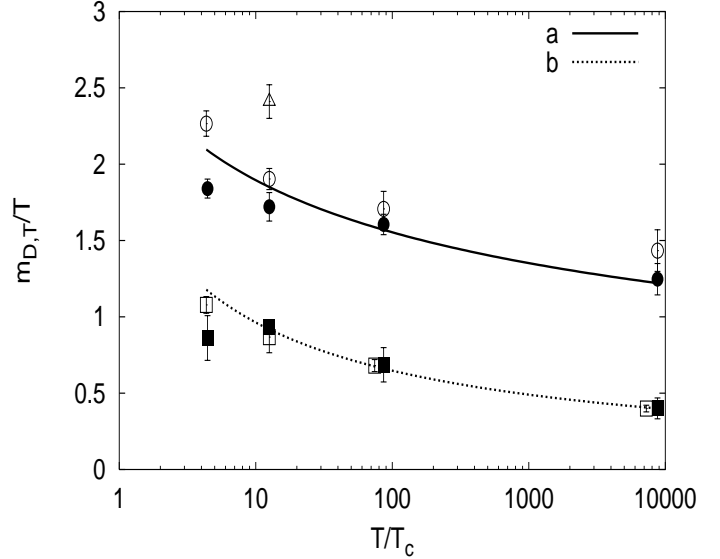}}
\caption{ The screening masses in units of the temperature.
Shown are the Debye mass
$m_D$ for the first (filled circles) and the second (open circles)
set of $h$, and
the magnetic mass  $m_T$ for the first (filled squares)
and the second (open squares) set of $h$.
The line (a) and line (b) represent fits to the temperature
dependence of the Debye and the magnetic mass from 4d
simulations of [6]. The magnetic mass for the set $II$
at temperature $T \sim 90 T_c$
and $ \sim 9000 T_c$ was shifted in the temperature scale for
better visualization. The open triangle is the value of the Debye mass
for $x=0.07$ and $h=-0.2179$.}
\end{figure}

Let us first discuss our calculations in the broken phase.
The simulations were done for two sets of parameters:
$\beta=16,~~x=0.03,~~h=-0.2181$ and $\beta=8~~,x=0.09,~~h=-0.5159$,
here $h$ was chosen along the perturbative line of 4d physics \cite{kajantie1}.
The propagators obtained by us in the broken phase show
a behaviour which is very different from
that in the symmetric phase and that in the 4d case studied in Ref.
\cite{heller1,heller2}. The magnetic mass extracted from the
gauge field propagators is $0.104(20)g_3^2$ for the first set 
and $0.094(8)g_3^2$ for the second set of the parameters.
It is by a factor 4 to 5 smaller than the corresponding 4d result.
Moreover, the propagator of the $A_0$ field does not seem to
show a simple exponential behaviour, what is in qualitative
agreement with the findings of Ref. \cite{rebhan1}.
These facts suggest that the broken phase can not
correspond to the physical phase.

We turn to the  discussion of our results in the symmetric phase.
In order to find the parameter range of interest for $h$ we first
have analyzed three different values of $h$ at $\beta=16$ and $x=0.07$.
The location of these values relative to
the transition line is shown in Figure~3. For the electric screening
mass we find, in increasing values of $h$,
$m_D/T = 1.72(10)$, $1.90(7)$ and $2.41(11)$.
 These results should be
compared to the 4d data. From the fit given in Ref.\cite{heller2}, we find at
$T/T_c= 12.57$ for the electric screening mass  $m_D/T = 1.85$. This
shows that our third value for $h$ clearly is inconsistent with the
4d result. 
From a linear interpolation between the results at the three different
values for $h$ we find the best matching value, i.e. a point on the
line of 4d physics, $h(x=0.07)=-0.2496$.

The temperature dependence of the screening masses
obtained in the symmetric phase for these two sets of
parameters, which stay close to the transition line, is shown in Figure~4.
Also given there is the result of the 4d simulations \cite{heller2}, $m_D^2/T^2
 = Ag^2(T)$,
with $A=1.70(2)$ for the electric mass and $m_T/T= C g^2(T)$, with
$C=0.456(6)$ for the magnetic mass. As can be seen both masses can
be described consistently with a unique choice of the coupling $h$ for
temperatures larger than $10T_c$. Although even at $T\simeq 4T_c$ we
find reasonable agreement with the 4d fits, we note that the accurate choice 
of $h$ becomes important and a
simultaneous matching of the electric and magnetic masses seems to be
difficult. For larger temperatures we find that the magnetic mass
shows little dependence on $h$ (in the narrow range we have analyzed)
and the determination of the correct choice of $h$ thus is mainly
controlled by the variation of the electric mass with $h$.

Let us summarize our findings for the screening masses in the
symmetric phase.  For $T \ge 10 T_c$ the screening masses can be
described very well in the effective theory. Suitable values of $h$ can be
found using the interpolation procedure outlined above for $x=0.07$.
This procedure can also be followed for $x=0.05$ and $0.03$. There
the 4d data are well matched for values corresponding to set
$I$ (see Figure 4), therefore the following $h$ values can be considered
as ones corresponding to 4d physics, $h(x=0.07)=-0.2496$,
$h(x=0.05)=-0.2365$ and $h(x=0.03)=-0.2085$. An interpolation between
these values gives the approximate line of 4d physics $h_{4d}(x)$.

\section{Gauge independence of the propagator pole mass}

In confining theories the propagator pole does not correspond to an
asymptotic state (unlike e.g. in $QED$) therefore there is no
physical reason for gauge independence of the pole mass.
However, the propagator pole was proven in the framework of perturbation
theiry to be gauge 
independent in the high temperature deconfined phase of 
$SU(N)$ theory \cite{kobes}.

We have investigated the gauge dependence of the effective masses numerically
using the so-called $\lambda$-gauges \cite{bernard}, defined by the gauge 
fixing condition
\be
\lambda \partial_3 A_3+\partial_2 A_2+
\partial_1 A_1=0.
\ee
Case $\lambda=1$ corresponds to the Landau gauge.
The possible gauge dependence of the pole mass has been checked by 
investigating in addition to the
Landau gauge propagators also the propagators for
$\lambda=0.5$ and $2.0$ for $\beta=16,~~x=0.03$ and $h=-0.2085$ on
a $32^2 \times 96$ lattice. The results of these measurements together
with the results obtained from Landau gauge propagators (on $32^2 \times 64$
lattice) are summarized in Table 1.
\vspace{-1cm}
\begin{center}
$$ ~\beta=16~~$$
\begin{tabular}{|l|l|l|l|}
\hline
$~~~\lambda~~~$ &$~~~0.5~~~~$ &$~~~1.0~~~$ &$~~~2.0~~~$\\
\hline
$~~~m_D/g_3^2~~~$ &$1.22(5)$ &$1.32(5)$ &$1.20(10)$\\
\hline
$~~~m_T/g_3^2~~~$ &$0.48(5)$ &$0.46(8)$  &$0.42(5)$\\
\hline
\end{tabular}
\end{center}   
TAB 1: {\small Numerical investigation of the 
dependence of the pole mass on the gauge parameter $\lambda$.}

\section{Free energy resummation}
In this final section we will use the screening masses calculated
before for the evaluation of the free energy density.
The partition function can be calculated in the
effective theory in the following way:
\be
Z=Z_{non-stat} \int D A_0 D A_i exp( -\int d^3 x L_{eff})
\ee
where $Z_{non-stat}$ is the contribution of the non-static modes to the 
partition function which was calculated in \cite{braat1} to 3-loop
level,
and $L_{eff}$ is the lagrangian of the effective theory given by (1).
Performing the integration over static electric fields ($A_0$) one obtains the
contribution of the electric scale ($g T$) to the partition function
$Z_{el}$. This step can be performed perturbatively because $A_0$
has a non-zero thermal mass. The resulting contribution was calculated in
\cite{braat1} yielding odd powers to the weak coupling expansion of $\ln Z$.
Integration over static magnetic fields yields the contribution of length
scale $g^2 T$ to the partition function and was calculated 
using numerical lattice simulation in \cite{karsch3d}. In this way the
weak coupling expansion can be obtained to ${\cal O}(g^6)$.
However, there are at least two things one may worry about in the outlined 
perturbative program : i)In the calculation of $Z_{el}$ the tree-level (from the point of
view of the effective theory) mass $m_{D0}$ was used. As we have seen both
lattice simulations and coupled gap equations yield a mass for the $A_0$ 
field, which is considerably larger than the tree-level result $m_{D0}$.
ii) For $g \sim 1$ the separation of electric ($g T$)
and magnetic scales ($g^2 T$) is not obvious and it is interesting to 
investigate the influence of the latter on the $A_0$ integration. This 
investigation is also motivated by the fact that the 
electric mass is rather sensitive to the magnetic scale.
To investigate the effect of using the 'exact' masses 
we will reorganize the perturbation
theory and perform a loop expansion, instead of the expansion in powers of $g$.
 This can be achieved by rewriting the
Lagrangian as 
\ba
&&
L_{eff}={1\over 4} F_{ij}^a F_{ij}^a + \hal m_D^2 A_0^a A_0^a+
\hal m_T^2 A_i^a A_i^a+\nonumber\\
&&
{1\over 4} \lambda_A {(A_0^a A_0^a)}^2+L^m+L^{(2)}_{ct}+L^{(3)}_{ct},
\ea
where 
\be
L^m=\hal (m_{D0}^2-m_D^2) A_0^a A_0^a-\hal m_T^2 A_i^a A_i^a
\ee
and
\ba
&&
L^{(2)}_{ct}={g_3^2 N (N^2-1) m_{D}^2\over 4 {(4 \pi)}^2 \epsilon}+
{g_3^2 N (N^2-1) m_{T}^2\over 16 {(4 \pi)}^2 \epsilon}\hspace*{-0.2cm}\\
&&
L^{(3)}_{ct}={g_3^2 N (N^2-1) (m_{D0}^2-m_{D}^2)\over 4 {(4 \pi)}^2 \epsilon}-
\nonumber \\
& &
\hspace*{1.2cm}{g_3^2 N (N^2-1) m_{T}^2\over 16 {(4 \pi)}^2 \epsilon}
\ea
are the 2- and the 3-loop counterterms 
to be treated as perturbations.
The values of $m_D$ and $m_T$ will be taken from the coupled gap equation 
using the BP scheme.
At 2-loop level the following diagrams will contribute to the free energy,
$-\ln Z_{el}/V,$
\vskip0.2truecm
\epsfbox{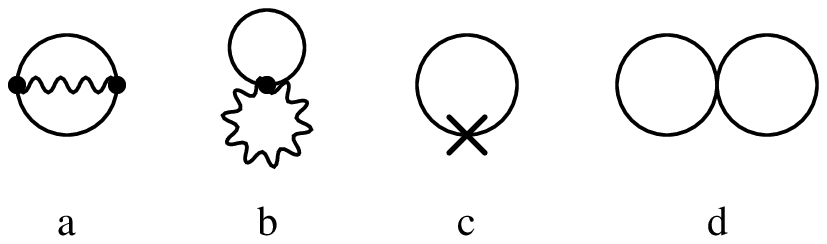}
\noindent
(The solid line denotes the $A_0$ propagator, the dot denotes
the $A_0-A_i$ vertex and the cross denotes the mass counterterm).
We will also study the 3-loop contribution in the limiting case
$m_T=0$. In this case the following 3-loop diagrams should be taken 
into account in addition to those calculated in \cite{braat1}:
\vskip0.2truecm
\epsfbox{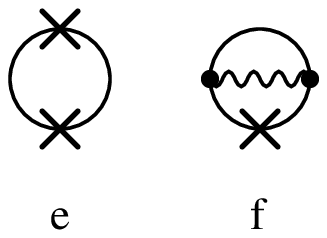}
\noindent
The contribution of these diagrams to $f_{el}=-\ln Z_{el}/V$ are:
\ba
&
f^a=-{(N^2-1) N g_3^2\over 4 {(4 \pi)}^2} ( 3 m_D^2 -\hal m_T^2-
\nonumber\\
&
2 m_D m_T+(4 m_D^2-m_T^2) \ln {\Lambda\over m_T+2 m_D})\\
&
f^b=-{(N^2-1) N g_3^2\over {(4 \pi)}^2}m_D m_T\\
&
f^c=-{(N^2-1) (m_{D0}^2-m_{D}^2) m_D\over 8 \pi}\\
&
f^d=-\lambda_A (N^4-1) {m_D^2\over {(4 \pi)}^2}\\
&
f^e=-{(N^2-1) {(m_{D0}^2-m_D^2)}^2\over 32 \pi m_D}\\
&
f^f={g_3^2 (N^2-1) N (m_{D0}^2-m_D^2)\over {(4 \pi)}^2} (\ln {\Lambda\over
2 m_D}+{1\over 4})
\ea
\begin{figure}
\epsfbox{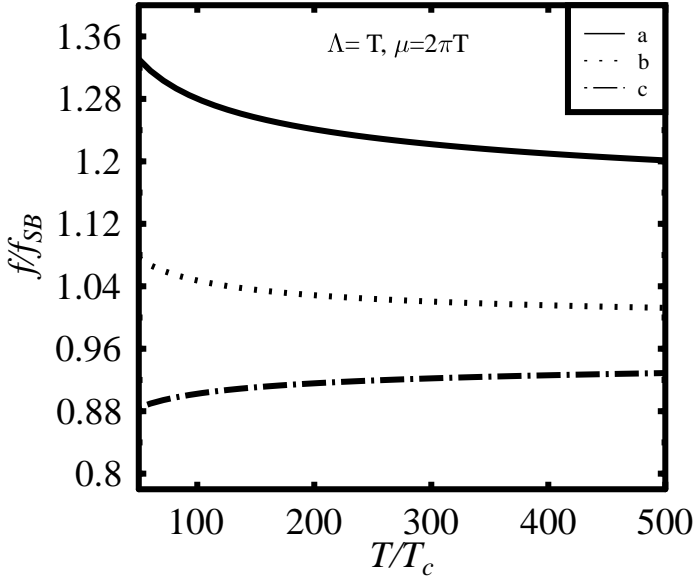}
\caption{The resummed loop expansion of the SU(3) free energy: a) the 1-loop,
b) 2-loop and c) the 3-loop level free energy for $m_T=0$.}
\end{figure}
\begin{figure}
\epsfbox{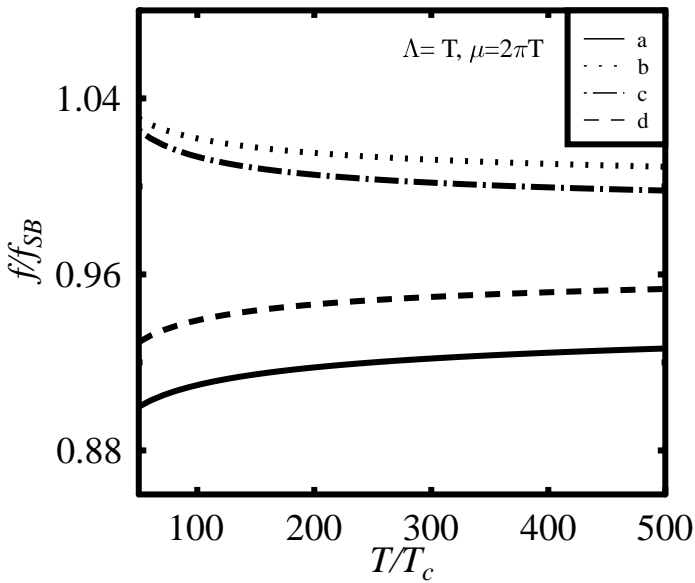}
\caption{The weak coupling expansion of the SU(3) free energy: 
a, b, c and d are the ${\cal O}(g^2)$, ${\cal O}(g^3)$, ${\cal O}(g^4)$
and ${\cal O}(g^5)$ level results}
\end{figure}
First let us investigate the resummed loop expansion for the free energy in
the case $m_T=0$. The numerical results of the loop expansion are shown in
Figure 5. For comparison we have also plotted in Figure 6 the weak coupling
expansion  for the free energy calculated in \cite{arnzhai,braat1} which
shows the known alternating behaviour in different orders. Compared to this
in the resummed loop expansion the contribution of the static modes is
larger, but the free energy decreases systematically, contrary to the 
alternating
behaviour of the weak coupling expansion. The 3-loop level resummed free
energy if compared to ${\cal O}(g^5)$ order result differ from it by the 
amount of a few percent. Finally the effect of massive magnetic modes was 
studied at 2-loop level and it turns out that the variation of the free energy
is about 1\% relative to the $m_{T}=0$ case.

\section{Summary}
The free energy density has been shown in this paper to be rather sensitive to
the actual values of the screening masses. This is probably also true for 
other physical quantities. This gives the motivation for the careful 
analytical and numerical investigation of these quantities, presented in our
paper.

Both numerical Monte-Carlo simulations and 
investigations of the coupled gap equations show that the magnetic mass
of the 3d adjoint Higgs model is very close to the magnetic mass of the 
pure 3d gauge theory, however, the numerical values of the magnetic mass
obtained in these aproaches are different. The 1-loop gap equation approach
gives the magnetic mass roughly equal to $0.28 g_3^2$ for BP scheme
and $0.38 g_3^2$ for AN scheme, while the numerical lattice simulations
give $0.46 g_3^2$.
The temperature dependence of the Debye mass was found very close to the 
leading order result both in the
gap equation approach and in the numerical lattice simulations, but the
numerical values are again different and equal to
$(1.2-1.3) m_{D0}$ for the coupled
gap equations ( for $T > 100 T_c$ ) depending on resummation scheme
(see Figure 1) and $1.6 m_{D0}$ for lattice simulations.

We leave the interpretation of these systematic deviations to a future
publication.
\vskip0.3truecm
\noindent {\bf Acknowledgements:}  
This work was partly supported by the TMR network {\it Finite Temperature
Phase Transitions in Particle Physics}, EU contract no. ERBFMRX-CT97-0122.
P.P. thanks V.P. Nair, O. Philipsen and K. Rummukainen for useful
discussions and the organizers of the TFT98 workshop for financial
support.

\end{document}